**Zero-field Nernst effect in a ferromagnetic kagome-lattice Weyl-semimetal $Co_3Sn_2S_2$**

*Satya N. Guin,\* Praveen Vir, Yang Zhang, Nitesh Kumar, Sarah J. Watzman, Chenguang Fu, Enke Liu, Kaustuv Manna, Walter Schnelle, Johannes Gooth, Chandra Shekhar, Yan Sun, and Claudia Felser\**

Dr. S. N. Guin, P. Vir, Y. Zhang, Dr. N. Kumar, Dr. C. Fu, Dr. E. K. Liu, Dr. K. Manna, Dr. W. Schnelle, Dr. J. Gooth, Dr. C. Shekhar, Dr. Y. Sun, and Prof. Dr. C. Felser
Max Planck Institute for Chemical Physics of Solids,
01187 Dresden, Germany
E-mails: satyanarayanguin@cpfs.mpg.de; Claudia.Felser@cpfs.mpg.de

Dr. S. J. Watzman
Department of Mechanical and Materials Engineering,
University of Cincinnati, Cincinnati OH 45221, USA

Dr. E. K. Liu
Institute of Physics, Chinese Academy of Sciences, Beijing 100190, China

**Abstract.** The discovery of magnetic topological semimetals recently attracted significant attention in the field of topology and thermoelectrics. In a thermoelectric device based on the Nernst geometry, an external magnet is required as an integral part. We report a zero-field Nernst effect in a newly discovered hard-ferromagnetic kagome-lattice Weyl-semimetal $Co_3Sn_2S_2$. A maximum Nernst thermopower of ~3 µV K$^{-1}$ at 80 K in zero field is achieved in this magnetic Weyl-semimetal. Our results demonstrate the possibility of application of topological hard magnetic semimetals for low-power thermoelectric devices based on the Nernst effect and are thus valuable for the comprehensive understanding of transport properties in this class of materials.

The advantages such as the absence of moving parts and noise make the thermoelectric technology promising for energy conversion and solid-state cooling.[1–6] Its progress largely relies on the Seebeck effect, i.e., the generation of an electrical voltage longitudinal to a temperature gradient. In contrast, configurations based on the Nernst effect, which is the generation of a transverse electrical signal in a magnetic field, have been significantly less studied,[7–9] partly as an external magnetic field is usually required to observe a Nernst signal. In conventional thermoelectric devices based on the Seebeck effect, the heat reservoir is required to be a part of the electrical circuit (**Figure 1**a). In contrast, the multiterminal thermoelectric devices based on the Nernst effect enable spatial separation of the heat reservoir from the electric circuitry.[10] Moreover, in Nernst devices, there is no need for both *p*- and *n*-type materials as the polarity of the voltage can be reversed by reversing the magnetic field direction.[8,11] Therefore, Nernst devices overcome certain problems of Seebeck devices, where the different thermal expansion coefficients of *p*- and *n*-type materials lead to compatibility issues.

In a ferromagnetic material, the generation of an additional electric voltage orthogonal to the applied temperature gradient due to the internal magnetization is referred to as the anomalous Nernst effect (ANE).[11,12] In general, the Nernst thermopower ($S_{xy}$) of a soft magnetic material is zero once the magnetic field is removed. Therefore, a part generating an external magnetic field needs to be integrated in devices based on the Nernst effect, which is a big obstacle for the thermomagnetic devices. A possible solution could be the use of a hard magnetic material, where an anomalous Nernst signal ($S_{xy}^A$) can be obtained in zero field. Such a permanent magnetic material remains in a magnetized state even when the external magnetic field is removed.[13,14] Magnetic ferrites and rare-earth-based hard magnets have been extensively studied owing to their wide range of functional applications including magnetic motors, magnetic recording media, magnetic fluids, and electromagnetic wave filters.[13–16] A novel potential application with hard magnets is energy conversion based on the Nernst effect, which has not attracted significant attention. Few examples of such materials include thin films of the alloys FePt, FePd, L1$_0$-MnGa, D0$_{22}$-Mn$_2$Ga, Co/Ni, dilute magnetic semiconductor (DMS) Ga$_{1-x}$Mn$_x$As, and recently Mn$_3$Sn.[17–20] Therefore, novel hard-magnetic materials are required. Studies on their thermomagnetic properties are of significance in terms of both technological application and fundamental understanding. A promising approach is to search for a new candidate from the library of topological materials as they share many mutual features with the thermoelectric materials.

In the current decade, topological materials attract significant research interest. These newly discovered materials exhibit various physical properties such as chiral anomaly, high carrier mobility, giant magnetoresistance, and mixed axial-gravitational anomaly due to the topological band structure.[21–26] They are also potential candidates for functional applications in quantum computing, infrared sensors, and heterogeneous catalysis.[27–29] In this context, the discovery of magnetic topological materials created a new frontier in the field.[30–37] These exotic magnetic topological materials provide a new scope in the thermoelectric research as the topological band structure effect can enhance the thermoelectric response.

In this communication, we show that the topological ferromagnets with large coercive fields ($H_c$) could be potential candidates for observation of the Nernst effect in zero field. As a case study material, we choose Co$_3$Sn$_2$S$_2$, a newly discovered hard-magnetic kagome-lattice Weyl-semimetal. Recently, a large intrinsic anomalous Hall conductivity (AHC) and giant anomalous Hall angle (AHA) have been observed in Co$_3$Sn$_2$S$_2$, originating from the topological band structure.[38,39] We show that a single-crystalline Co$_3$Sn$_2$S$_2$ exhibits a maximum $S_{xy}^A$ value of ~3 µV K$^{-1}$ at 80 K in zero field, which is significantly high considering its magnetic moment of ~0.89 µ$_B$ f.u.$^{-1}$. We employed a combined approach including electrical and thermoelectric measurements and first-principle calculations to elucidate this observation. Our complementary electrical transport measurement and density-functional-theory (DFT)-based calculations indicate that the high Nernst signal in Co$_3$Sn$_2$S$_2$ originates from the topological band structure.

The ternary chalcogenide $Co_3Sn_2S_2$ is a member of the shandite family $A_3M_2X_2$ ($A$ = Ni, Co, Rh, Pd; $M$ = Pb, In, Sn, Tl; $X$ = S, Se) and crystallizes in a rhombohedral structure (space group: $R\bar{3}m$).[40,41] In this structure, Sn atoms are distributed over interlayers (Sn1) and in the kagome layers (Sn2). The Sn(1) atoms are surrounded by two $Co_3$ triangles in a trigonal-antiprismatic configuration, whereas the Sn(2) atoms form a hexagonal planar layer with the magnetic Co atoms arranged on a kagome lattice in the $ab$-plane (Figure 1c). Each of the $Co_3$ triangles is capped above or below the kagome sheets by an S atom. The material is a type-IA half-metallic ferromagnet with a Curie temperature of $T_c$ = 177 K and magnetic moment of 0.29 $\mu_B$/Co.[40,41] The strong magnetic anisotropy leads to a long-range quasi-two-dimensional (2D) type of magnetism with a spontaneous out-of-plane magnetization.[40,41]

The single-crystals of $Co_3Sn_2S_2$ were synthesized from the elements using a melting reaction followed by slow cooling. The as-grown crystal was characterized by Laue X-ray diffraction (XRD) and powder XRD (see Methods and Figure S1, Supporting Information). For the electrical and thermoelectric measurements, the crystal was oriented and cut into a bar shape. As the material exhibits a highly anisotropic out-of-plane magnetization, we used the following configuration: magnetic field ($\mu_0H$) || $c$-axis and electrical current ($I$) or thermal gradient ($\Delta T$) || $ab$-plane.

Figure 1d shows the magnetization ($M$) as a function of the magnetic field ($\mu_0H$) for $Co_3Sn_2S_2$ at different temperatures. Distinct rectangular-shaped hysteresis loops with large coercive fields $H_c$ of ~3650 and ~570 Oe are observed at 2 and 100 K, respectively. The origin of the large coercivity is attributed to the strong magnetic anisotropy of the compound (see discussion, Supporting Information).[42] This distinct hysterics loop is observed up to ~100 K; it becomes less pronounced with the further increase in temperature and vanishes as the temperature approaches $T_c$. The temperature-dependent magnetization measurement indicates a magnetic transition at ~177 K (Figure S2, Supporting Information). Overall, the magnetic measurement data indicate that $Co_3Sn_2S_2$ exhibits a sizable $H_c$ up to ~120 K.

After we demonstrated the hard-magnetic nature of $Co_3Sn_2S_2$, we investigated the anomalous Nernst thermopower ($S_{xy}^A$) in zero magnetic field. We magnetized the sample by applying a field of ±1 T (> $H_c$) to orient the magnetic moments along the *c*-axis and then switched off the field. **Figure 2**a presents $S_{xy}^A$ of $Co_3Sn_2S_2$ in zero field as a function of the temperature. $S_{xy}^A$ increases with the temperature and reaches a peak value of ~3 µV K$^{−1}$ at ~80 K. The further increase in the temperature leads to a decrease in $S_{xy}^A$; above ~120 K, the signal is almost zero as the compound loses its hard-magnetic structure. The observed maximum value of $S_{xy}^A$ for $Co_3Sn_2S_2$ is remarkable considering the magnetic moment of ~0.89 µ$_B$ f.u.$^{-1}$ (see discussion in Supporting Information). Further, we investigated the magnetic-field dependence of the Nernst thermopower $S_{xy}(\mu_oH)$ at different temperatures (Figure 2b and S4). As discussed in the introduction that ferromagnets exhibit ANE below their $T_C$, this system starts to exhibit an anomalous behavior in $S_{xy}$. The effect of the large magnetocrystalline anisotropy of this compound is evident in the $S_{xy}(\mu_oH)$ data. The anomalous behavior in $S_{xy}(\mu_oH)$ (Figure 2b and S4) can be observed up to the Curie temperature of the compound ($T_C$ = 177 K). Above $T_C$ the $Co_3Sn_2S_2$ do not show any anomalous Nernst effect as it is no longer magnetic in nature. Below 100 K, a rectangular-shaped hysteresis loop was observed in the $S_{xy}(\mu_oH)$ data; $S_{xy}$ maintains a plateau value after a certain field. The change of the sign of $S_{xy}$ is due to the flipping of the magnetic moment of the Co atoms with the direction of $\mu_oH$ (positive to negative). This observation is consistent with the zero-field measurement of a magnetized sample. The finite $H_c$ in the data below ~120 K enables to estimate the zero-field $S_{xy}$, i.e., $S_{xy}^A$. The temperature-dependent estimated values of $S_{xy}^A$ from the field sweep data are in good agreement with the temperature-dependent zero-field measurement results (inset in Figure 2b). In the $S_{xy}(\mu_oH)$ data, we observed $S_{xy}^A$ up to ~170 K by extrapolating the slope of the high-field data; however, estimated $S_{xy}^A$ above ~120 K is not a *true* zero-field anomalous value, as the hard-magnetic nature disappears above ~120 K.

In order to understand the origin of the large $S_{xy}^A$ of $Co_3Sn_2S_2$, we measured the electrical resistivity ($\rho_{xx}$), longitudinal thermoelectric response, i.e., the Seebeck coefficient ($S_{xx}$), and Hall resistivity ($\rho_{yx}$). The temperature dependence of the electrical resistivity ($\rho_{xx}$) is presented in **Figure 3**a. $\rho_{xx}$ decreases with the decrease in the temperature reaching a value of ~55 μΩ cm at 2 K, indicating the metallic nature of the compound. The anomaly at $T_c$ = 177 K in $\rho_{xx}$ reflects the onset of a magnetic transition. The negative sign of $S_{xx}$ is attributed to the dominant *n*-type charge carriers in $Co_3Sn_2S_2$ (Figure 3b). $S_{xx}$ linearly increases with the temperature and becomes approximately constant in the range of 50 to 120 K. Above 120 K, $S_{xx}$ increases with the temperature and exhibits an anomaly near $T_c$.

It is known that the intrinsic band structure effect from the Berry curvature can lead to a large anomalous Hall effect (AHE) in a topological material (see Methods). Therefore, we estimated the Hall conductivity ($\sigma_{xy}$) of $Co_3Sn_2S_2$ from the measured electrical resistivity ($\rho_{xx}$) and Hall resistivity ($\rho_{yx}$):

$$\sigma_{xy} = \frac{\rho_{yx}}{\rho_{yx}^2 + \rho_{xx}^2} \qquad (1)$$

Figure 3c presents $\sigma_{xy}$ of $Co_3Sn_2S_2$ as a function of the magnetic field at different temperatures. The trend and values of the AHC ($\sigma_H^A$) are fully consistent with previous observations.[38] Below 100 K, a large AHE is observed with a sharp rectangular-shaped loop, which arises from the topological-band-structure-enhanced Berry curvature effect.[38] This observation indicates the topological-band-structure-related origin of the large $S_{xy}^A$ of $Co_3Sn_2S_2$ as the transverse thermoelectric response is associated with the Berry curvature, as discussed below.

The notion of the topological band structure effect in transverse thermoelectric responses can be understood by analyzing the temperature dependence of the anomalous transverse thermoelectric conductivity ($\alpha_N^A$). Therefore, we estimated the transverse thermoelectric conductivity ($\alpha_{yx}$) using the measured components of resistivity ($\rho_{xx}$, $\rho_{yx}$) and Seebeck and Nernst thermopowers ($S_{xx}$ and $S_{xy} = -S_{yx}$):

$$\alpha_{yx} = \frac{S_{yx}\rho_{xx} - S_{xx}\rho_{yx}}{\rho_{xx}^2 + \rho_{yx}^2} \quad (2)$$

Figure 3d shows the anomalous part of $\alpha_{yx}$ of $Co_3Sn_2S_2$. The $S_{xy}$ and $\rho_{yx}$ exhibit anomalous behavior below $T_C$. Therefore, a non-zero $\alpha_N^A$ value can be observed only below $T_C$. In $Co_3Sn_2S_2$, the $\alpha_N^A$ shows a peak value around 150 K and decreases with upon cooling, although the saturation magnetization increases. A similar observation was reported in an anomalous Hall resistance measurement.[38] The observation is an evidence that the anomalous Nernst signal originates from a different source, rather than from the magnetization.

For a deeper understanding, we carried out DFT- and *ab-initio*-based electronic structure calculations and numerical simulations of $\sigma_H^A$ and $\alpha_N^A$ using the Kubo formula. **Figure 4**a shows the calculated electronic structures of $Co_3Sn_2S_2$ with and without spin–orbit coupling (SOC). The spin-down channel of the bands has a gap of ~0.35 eV, while the spin-up channel is semimetallic (see also Figure S6). Moreover, for the spin-up channel along the Γ–L and L–U paths, linear band crossings can be observed near the charge-neutral Fermi point ($E_0$). In the absence of SOC, the full Hamiltonian could be split into the direct product of a spinless Hamiltonian and spin-Hamiltonian, in which the band degeneracy of each spin channel is determined by the spinless Hamiltonian. Owing to the mirror symmetry of the lattice, the band inversion of the spin-up channel forms a nodal line in the mirror plane ($M_y$). Considering the combination of inversion and $C_3$ rotational symmetries, there are six nodal lines in total (Figure 4b). The application of SOC breaks the mirror symmetry and lifts the degeneracy of the nodal lines, with one pair of Weyl points remaining on it.

Figure 4c shows the computed $\alpha_N^A$ as a function of the position of the Fermi level ($E_F$) at 80 K. We would like to note that unlike intrinsic AHE, the intrinsic part of $\alpha_N^A$ depends on the strength of the Berry curvature near $E_F$ rather than on the Berry curvature of all of the occupied bands (see Methods).[43,44] The anomalous Nernst conductivity (ANC) of $Co_3Sn_2S_2$ exhibits a strong $E_F$ dependence, which is not observed for $\sigma_H^A$ (Figure S7). $\sigma_H^A$ maintains a plateau value (~1000 S cm$^{-1}$) over an energy window of ~100 meV around $E_F$. In contrast, $\alpha_N^A$ has a peak value slightly away from $E_F$ and then rapidly decreases away from $E_F$ due to the modulation of the Berry curvature strength. This finding was further supported by our temperature-dependent $\sigma_H^A$ and $\alpha_N^A$ data. The experimental $\sigma_H^A$ exhibits a small change below ~100 K. In contrast, $\alpha_N^A$ exhibits a monotonous decreasing trend with the decrease in temperature.

In a previous study, angle-resolved photoelectron spectroscopy and low-temperature Shubnikov–de Haas (SdH) quantum oscillation analysis indicated that $E_F$ was slighted shifted toward the valence bands in $Co_3Sn_2S_2$.[38] In this study, we use a sample with a similar quality to that for the AHE, indicating a similar position of $E_F$. We calculated the temperature dependence of $\alpha_N^A$ below ordering temperature using different positions of $E_F$ of $Co_3Sn_2S_2$ (Figure S8). $\alpha_N^A$ exhibits a monotonous increase with $T$ for $E_F = E_0 - 0.07$ eV to $E_0 - 0.10$ eV and well agrees with the experimental value when $E_F - E_0 \sim -0.08$ eV (Figure 4d). Moreover, the temperature-modified Berry curvature distributions ($\Omega_N$) at 80 and 150 K for $E_F - E_0 \sim -0.08$ eV also have large nonzero values (Figure 4b and S9), suggesting a topological band structure contribution in the transverse thermoelectric transport.

**Figure 5** compares the absolute value of peak $S_{xy}^A$ of different class of materials from the literature with $Co_3Sn_2S_2$.[17–19,34,45–49]. A maximum $S_{xy}^A$ of ~3 μV K$^{-1}$ at $T = 80$ K is measured in the case of $Co_3Sn_2S_2$. In addition, $Co_3Sn_2S_2$ exhibits a zero-field signal, which is not the case for nonmagnetic and soft- ferromagnetic metals. Although the maximum $S_{xy}^A$ of $Co_3Sn_2S_2$ (~3 μV K$^{-1}$ at $T = 80$ K) is lower than that of topological ferromaget $Co_2MnGa$ (~6 μV K$^{-1}$ at $T = $ 300 K) and DMS $Ga_{0.93}Mn_{0.07}As$ (~8.1 μV K$^{-1}$ at $T = 10$ K), but the present results provides a possible guiding principle for the observation of large zero-field anomalous Nernst thermopower in a hard-ferromagnetic topological semi-metal at elevated temperature. Moreover, considering the magnetic moments of ~0.89 $\mu_B$ f.u.$^{-1}$ for $Co_3Sn_2S_2$ ($T$ = 80 K) and ~3.7 $\mu_B$ f.u.$^{-1}$ for $Co_2MnGa$ ($T$ = 300 K), the present result is more interesting and due to lower magnetic moment the inherent stray field for $Co_3Sn_2S_2$ will be low. Figure S10 compares the ratios of the strengths of the anomalous Nernst signals to their saturation magnetizations of different hard and soft ferromagnetic metals and antiferromagnet $Mn_3Sn$.[17,19,34,45–49] The Y-axis of the plot ($S_{xy}^A/\mu_0 M$) represents the ratio of $S_{xy}^A$ to the magnetic moment ($\mu_0 M$). The values of $S_{xy}^A/\mu_0 M$ for the topological ferromagnets are higher than those of the trivial ferromagnetic materials. A similar result was observed for the topological chiral antiferromagnet $Mn_3Sn$; however, it exhibited a lower Nernst signal than that of $Co_3Sn_2S_2$ (Figure S10).[19] $S_{xy}^A/\mu_0 M$ indicates that the strength of the transverse signal is significantly higher in $Co_3Sn_2S_2$ and indeed highest among those of ferromagnetic metals. This implies that the anomalous transport properties originate from the topological band structure effect and are independent of the strength of the magnetic moments.

In conclusion, magnetic topological semimetals are potential candidates for the observation of exotic anomalous transport properties. We showed that the hard ferromagnetic kagome-lattice Weyl-semimetal $Co_3Sn_2S_2$ from the shandite family exhibited a large ANE in zero field. Our experimental and DFT calculation results showed that the large ANE originated from the strong Berry curvature associated with the nodal lines and Weyl points. Although the magnitude of the zero-field transverse signal of $Co_3Sn_2S_2$ was not sufficiently high for applications, this study shows that the topological material with a large $H_c$ could be a potential candidate for a transverse thermoelectric and that the large signal can be achieved in zero field. Therefore, it is of interest to search for new hard magnets with high Curie temperatures from the library of magnetic topological materials.

## Experimental Section

### Single-crystal growth of $Co_3Sn_2S_2$ and characterization

The single-crystals of $Co_3Sn_2S_2$ were grown using an elemental melting reaction followed by a slow cooling. As-purchased high-quality elemental cobalt (99.999%), tin (99.999%), and sulphur (99.999%) were used for the synthesis. Stoichiometric amounts of the elements (Co:Sn:S = 3:2:2) were loaded in an alumina crucible and then sealed in a quartz tube. The tube was heated to 1323 K over a period of 48 h, soaked for 24 h, and then slowly cooled down to 873 K over 7 days. The power XRD measurement was performed at room temperature using a Huber Image Plate Guinier Camera G670 operated with Cu $K_{\alpha 1}$ radiation ($\lambda = 1.54056$ Å). The single-crystallinity of the as-grown crystal was evaluated by the white-beam backscattering Laue XRD method. Well-characterized and aligned crystals were cut into bar shapes for transport and magnetization measurements. The typical dimensions of the crystals used for the electrical and thermal transport measurements were ~$7 \times 1.3 \times 0.4$ mm$^3$.

### Electrical transport and magnetization measurements

The electrical transport properties were measured using a PPMS9 instrument (ACT option, Quantum Design). The standard four-probe method was used in all of the measurements. In order to correct for contact misalignment, the measured data were field-symmetrized and antisymmetrized. The magnetization measurement was performed using a Quantum Design MPMS3 instrument.

### Thermoelectric transport measurements

All of the thermal transport experiments were performed in a physical property measurement system (PPMS) cryostat. The Seebeck and Nernst thermoelectric measurements were carried out in the one-heater two-thermometer configuration. In the field sweep experiments in a temperature range of 13–310 K, the PPMS as well as an external nanovoltmeter and current source (Keithley) were controlled using LabVIEW. The temperature gradient was generated using a resistive heater, connected to a gold-coated flat copper wire at one end of the sample.

The thermal gradient $\Delta T$ was applied along the *ab*-plane of the sample, while the magnetic field was applied along the *c*-axis. Another gold-plated flat copper wire was attached to the puck clamp for the heat sink. In order to measure the temperature gradient, two gold-plated copper leads were attached directly to the sample using a silver-filled epoxy along the thermal gradient direction. The distance between the thermometers was ~3.5 mm. In the Seebeck and Nernst measurements, $\Delta T$ was typically set to ~1–3% of the base temperature. Two copper wires were attached using the silver epoxy, orthogonal to the thermal gradient direction, to measure the transverse voltage. In order to correct the data for contact misalignment, the measured data were field-symmetrized and antisymmetrized. The thermal conductivity measurement was carried out in the PPMS using the thermal transport option.

**Anomalous Nernst thermopower of different class of materials form the literature**

We use the highest anomalous Nernst thermopowers ($S_N^A$) for the different soft and hard ferromagnets, antiferromagnet Mn$_3$Sn and dilute magnetic semiconductor Ga$_{0.93}$Mn$_{0.07}$As well below their Curie temperatures. We used their magnetizations ($\mu_0 M$ in Tesla) to calculate the ratios. (Co/Ni film (300 K; ref. 17), L1$_0$–FePt (300 K; ref. 17), D0$_{22}$–Mn$_2$Ga (300 K; ref. 17), L1$_0$–MnGa (300 K; ref. 17), L1$_0$–FePd (300 K; ref. 17), Ga$_{0.93}$Mn$_{0.07}$As (10 K, ref. 18), Mn$_3$Sn (200 K; ref. 19), Co$_2$MnGa (300 K; ref. 34, 36), Nd$_2$Mo$_2$O$_7$ ($T < T_c = 73$ K, $B = 1$ T [111]; ref. 45), Fe (300 K; ref. 46), Co (300 K; ref. 46), Fe$_3$O$_4$ (300 K, $B < 0.8$ T; ref. 47), MnGe (100 K, $B > 5$ T; ref. 48), and Pt/Fe multilayer ($N = 9$, 300 K, $B < 5$ T; ref. 49)).

***ab-initio* calculations**

For our *ab-initio* calculations, we employed the DFT using VASP,[50] including the exchange–correlation energy through the Perdew–Burke–Ernzerhof functional. For the integrations in the *k*-space, we used a grid of $19 \times 19 \times 19$ points. We extracted the Wannier functions from the resulting band structure using WANNIER90[51] to set up a tight-binding Hamiltonian, which reproduced the DFT band structure within a few millielectronvolts. With this Hamiltonian, we calculated the Berry curvature $\vec{\Omega}_n$:

$$\Omega_{n,ij} = Im \sum_{m \neq n} \frac{\langle n|\frac{\partial H}{\partial k_i}|m\rangle \langle m|\frac{\partial H}{\partial k_j}|n\rangle - (i \leftrightarrow j)}{(\varepsilon_n - \varepsilon_m)^2}$$

where $m$ and $n$ are the eigenstates and $\varepsilon$ are the eigenenergies of the Hamiltonian $H$. Subsequently, we calculated the AHC $\sigma_H^A$:

$$\sigma_H^A = -\frac{e^2}{\hbar} \sum_n \int \frac{dk}{(2\pi)^3} \Omega_{n,xy}(k) f_{nk}$$

where $f_{nk}$ is the Fermi–Dirac distribution for a band $n$ at a $k$-point. The ANC $\alpha_N^A$ below ordering temperature is:[43]

$$\alpha_N^A = \frac{e}{T\hbar} \int \frac{dk}{(2\pi)^3} \Omega_N$$

If we define $\beta = k_B T$, where $k_B$ is the Boltzmann constant, the temperature-modified Berry curvature distribution can be expressed as:

$$\Omega_N = \sum_n \Omega_{n,yx}(k)\{(\varepsilon_{nk} - \mu)f_{nk} + k_B T \ln[1 + e^{-\beta(\varepsilon_{nk}-\mu)}]\}$$

$\Omega_N$ has a finite value only around the Fermi energy. Therefore, $\sigma_H^A$ is related to the summation of the Berry curvatures of all of the occupied bands below the Fermi energy. In contrast, $\alpha_H^A$ is related to the Berry curvature of the occupied bands near the Fermi energy. For the integrations over the whole Brillouin zone in the AHE and ANE calculations, we employed a mesh of 251 × 251 × 251 $k$-points, providing converged results. Discussion on finite temperature DFT calculation has been given in supporting information.

**Supporting Information**

Supporting Information is available from the Wiley Online Library or from the author.

**Acknowledgements**

This study was financially supported by the ERC Advanced Grant No. (742068) "TOP-MAT". S.N.G., Ch.F., and E.K.L. thank the Alexander von Humboldt Foundation for the fellowships. E.K.L. also thanks the National Natural Science Foundation of China (No. 51722106).

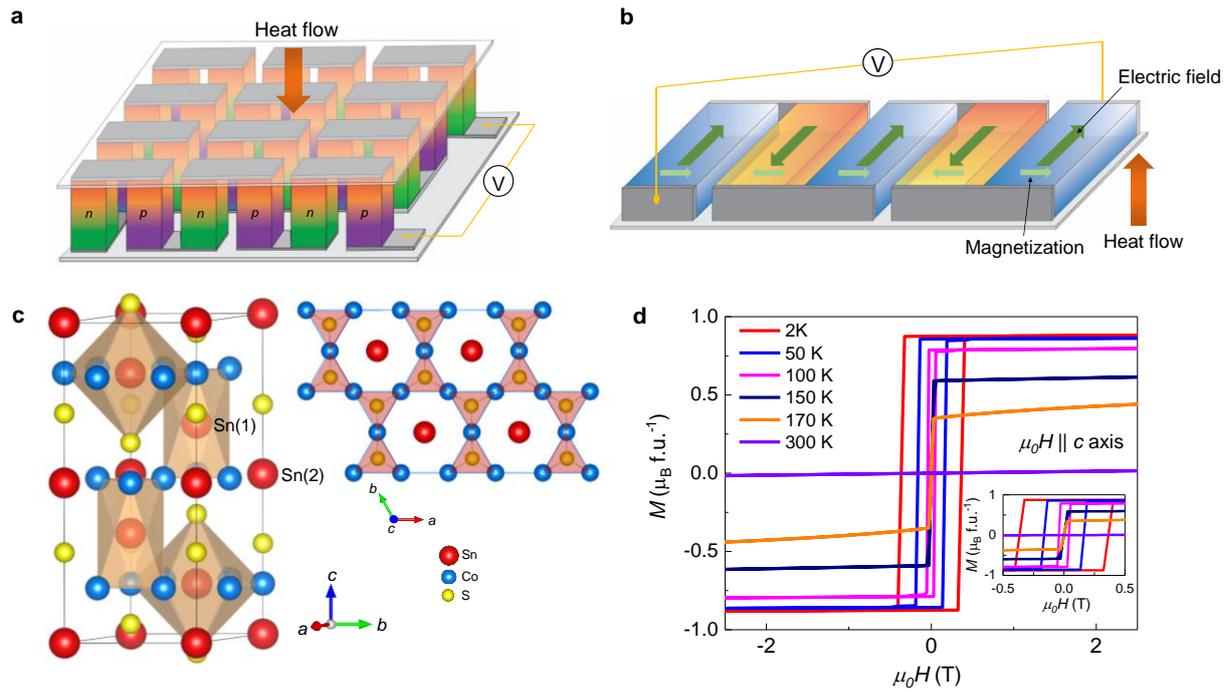

**Figure 1.** Schematics of thermoelectric modules, crystal structure, and magnetization of $Co_3Sn_2S_2$. a) and b) Thermoelectric modules in the Seebeck and Nernst geometries, respectively. In the conventional Seebeck device, both *n*- and *p*-type legs are required. In contrast, in the Nernst device, a single type of thermopile can be used and the polarity of the voltage can be altered by changing the magnetization direction. c) Crystal structure of $Co_3Sn_2S_2$. Sn atoms are distributed over the interlayer (marked as Sn(1)) and in the kagome layers (marked as Sn(2)). The structure consists of the flat hexagonal kagome unit composed of Co and Sn(2) (see the *c*-axis view). The Sn(1) atoms are surrounded by a couple of $Co_3$ triangles in a trigonal-antiprismatic configuration. S atoms capped the $Co_3$ triangles above or below the kagome sheets. d) Field-dependent magnetization of $Co_3Sn_2S_2$ at different temperatures. The inset shows a magnified plot.

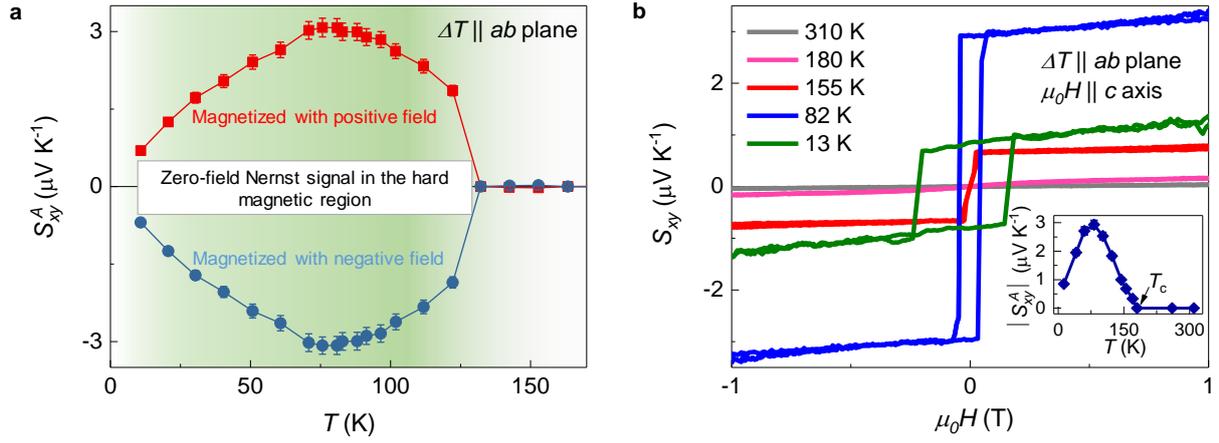

**Figure 2.** Nernst thermopower data. a) Temperature-($T$)-dependent anomalous Nernst thermopower ($S_{xy}^A$) of $Co_3Sn_2S_2$ in zero field, measured using a magnetized sample. b) Magnetic field dependences of the Nernst thermopower ($S_{xy}$) of $Co_3Sn_2S_2$ at different temperatures. The extracted zero-field anomalous Nernst thermopower ($S_{xy}^A$) as a function of $T$ is presented in the inset.

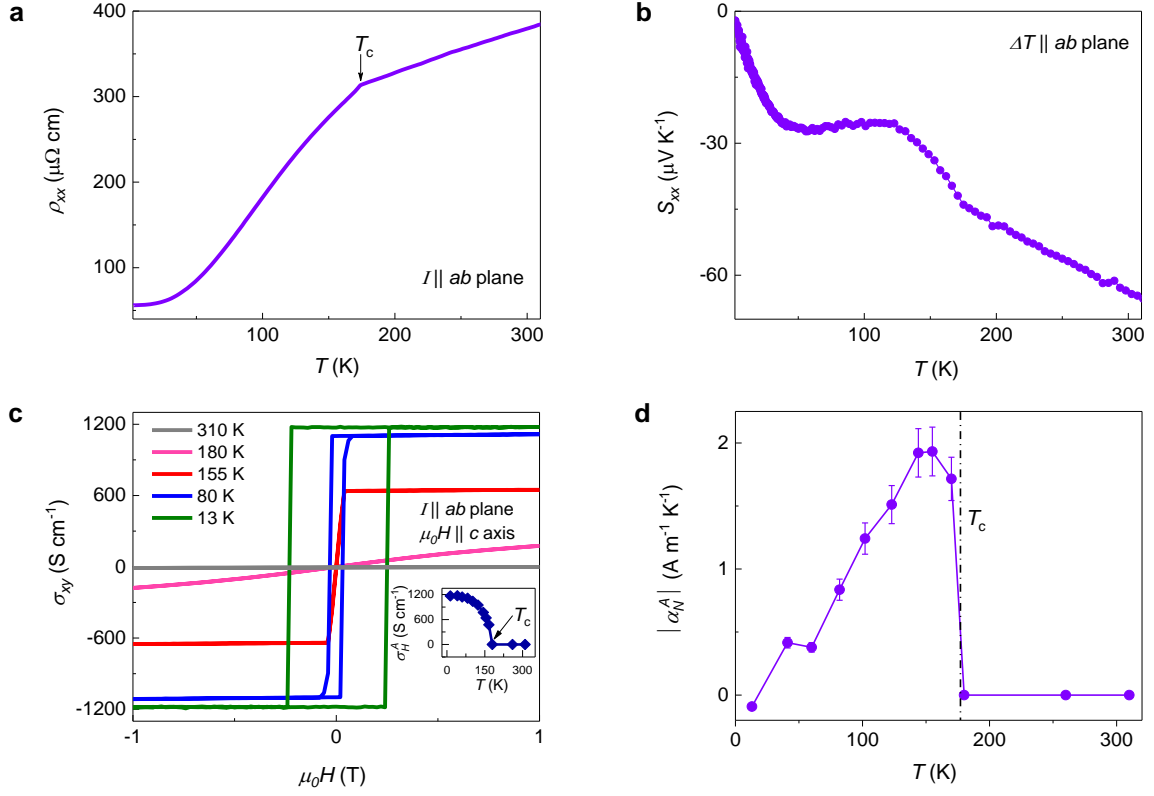

**Figure 3.** Electrical transport, Seebeck coefficient, and thermoelectric conductivity. a) $T$-dependent electrical resistivity ($\rho_{xx}$). The change in slope in $\rho_{xx}$ around ~177 K corresponds to a magnetic transition (marked with an arrow). b) $T$ dependence of the Seebeck coefficient ($S_{xx}$) of $Co_3Sn_2S_2$. c) Magnetic field dependence of the Hall conductivity ($\sigma_{xy}$). The $T$ dependence of the AHC ($\sigma_H^A$) is shown in the inset. d) $T$ dependence of the experimentally determined anomalous transverse thermoelectric conductivity ($\alpha_N^A$). The dotted line indicates magnetic transition temperature.

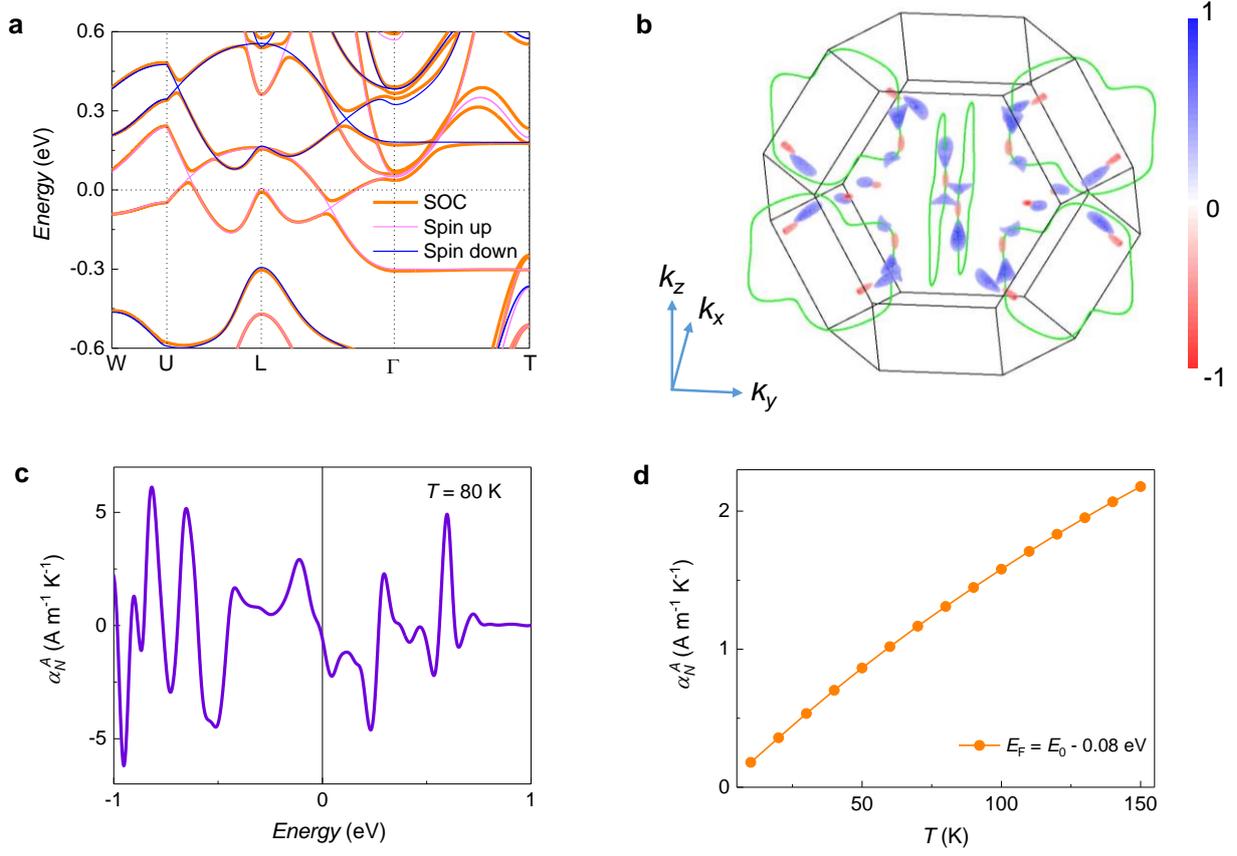

**Figure 4.** Electronic structure and calculated transverse thermoelectric conductivity. a) Electronic structures of $Co_3Sn_2S_2$ with and without SOC. The application of SOC breaks the mirror symmetry leading to gapping of the nodal lines due to the degeneracy lifting of the bands. b) Temperature-modified Berry curvature distribution ($\Omega_N$) at 80 K in the Brillouin zone with the magnetization along the $c$-axis for $E_F = E_0 - 0.08$ eV. The green lines indicate the nodal rings. c) Theoretically calculated ANC ($\alpha_N^A$) as a function of the chemical potential with respect to the charge neutrality point $E_0$ at $T = 80$ K. d) Theoretically estimated temperature-dependent $\alpha_N^A$ with $E_F = E_0 - 0.08$ eV below ordering temperature. The calculated value is in good agreement with the measured value.

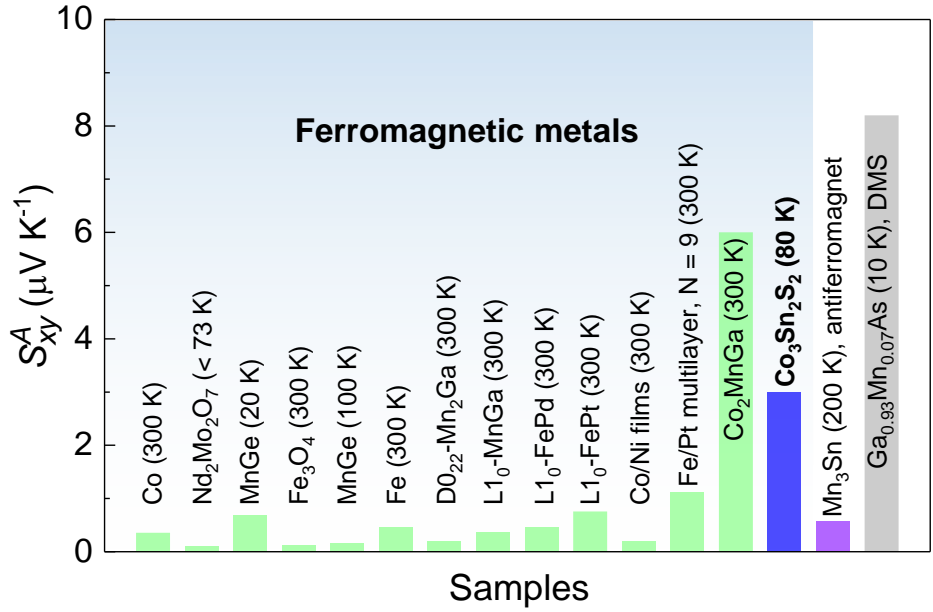

**Figure 5.** Absolute value of anomalous Nernst thermopowers ($S_{xy}^A$) for ferromagnetic metals, antiferromagnet $Mn_3Sn$, dilute magnetic semiconductor (DMS) $Ga_{0.93}Mn_{0.07}As$ from the literature and $Co_3Sn_2S_2$. The shaded region indicates hard and soft ferromagnetic metals.[17–19,34,45–49]

# Supporting Information

**Zero-field Nernst effect in a ferromagnetic kagome-lattice Weyl-semimetal $Co_3Sn_2S_2$**

Satya N. Guin,* Praveen Vir, Yang Zhang, Nitesh Kumar, Sarah J. Watzman, Chenguang Fu, Enke Liu, Kaustuv Manna, Walter Schnelle, Johannes Gooth, Chandra Shekhar, Yan Sun, and Claudia Felser*

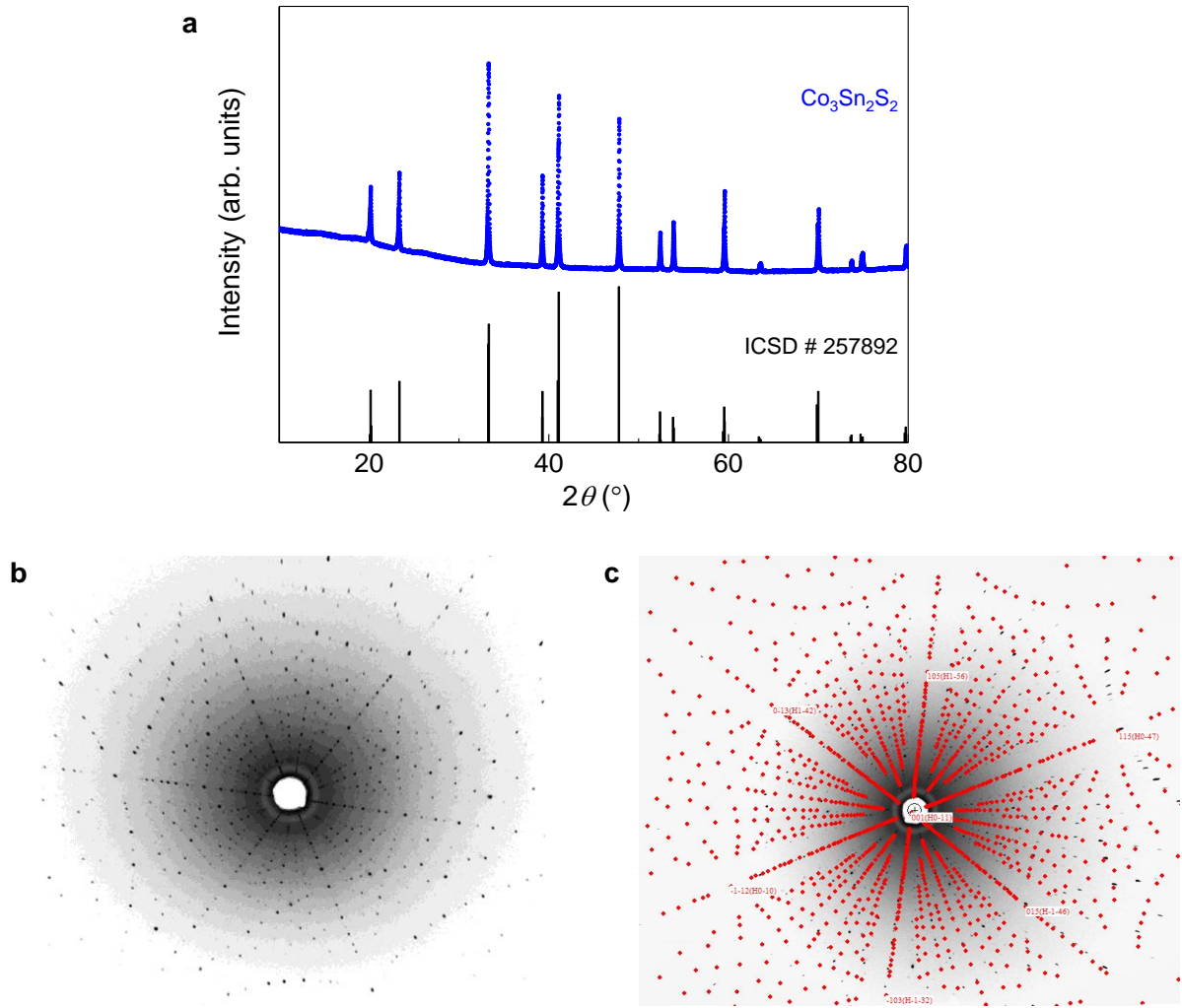

**Figure S1.** Characterization of the Co$_3$Sn$_2$S$_2$ crystal. a) Powder XRD pattern (Cu $K_{\alpha 1}$; $\lambda$ = 1.54056 Å) of a crashed crystal and corresponding simulated pattern. The pattern can be indexed based on the $R\bar{3}m$ space group. b) Laue diffraction pattern of the as-grown crystal. The diffraction pattern of the crystal can be indexed based on the $R\bar{3}m$ space group. c) Laue diffraction pattern of the [0001]-(*c*-axis)-oriented crystal and overlaid theoretically simulated pattern. The oriented crystal was cut into a bar shape using a wire saw for the transport and magnetization measurements.

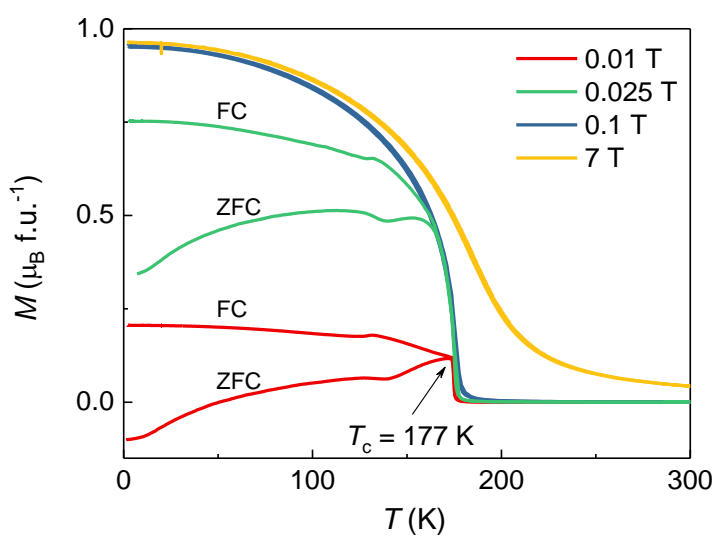

**Figure S2.** Magnetic measurements. Temperature-dependent zero-field-cooled (ZFC) and field-cooled (FC) magnetizations. The black arrow in the figure indicates the Curie temperature ($T_c$ = 177 K). A small change in magnetization is observed at ~140 K, which is related to the anomalous magnetic transition, consistent with the previous report.[S1]

**Origin of the large coercivity**

A [119]Sn Mossbauer spectroscopy indicated that Sn was strongly involved in the magnetic coupling within and between the Co layers. The electron correlations in the Co–S bond are reduced after the change to Co–Se by the substitution of S with Se. Therefore, the spin flipping is easier as the coupling between the Co layers is weakened, which is evident from the decreases in the hysteresis and $T_c$. This indicates that the strong magnetic anisotropy in $Co_3Sn_2S_2$ is the origin of the large coercivity.[S1–S3]

**Measurement geometry and definition of parameters**

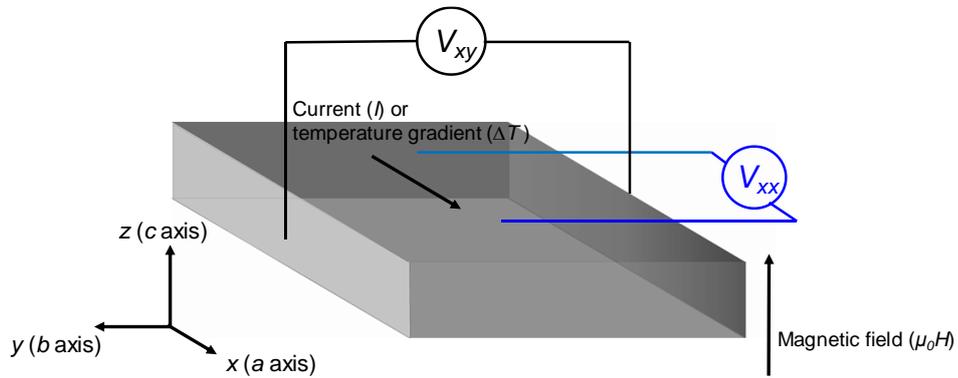

**Figure S3.** Schematic of the measurement geometry. The thermal gradient ($\Delta T$) and current ($I$) were applied along the *ab*-plane. The magnetic field ($\mu_0 H$) was applied parallel to the *c*-axis.

In order to clarify the notation of transverse and longitudinal responses (electrical and thermoelectric), in the below table, we define all of the transport parameters with their corresponding directions used in this study. A more detailed discussion on the transport coefficients is presented in ref. S4.

**Table S1**. Notation of the transport parameters.

| Symbol | Transport parameter | Symbol | Transport parameter |
|---|---|---|---|
| $S_{xx}$ | Seebeck thermopower ($\mu$V K$^{-1}$) | $S_{xy}$ | Nernst thermopower ($\mu$V K$^{-1}$) |
| $\rho_{xx}$ | Electrical resistivity ($\Omega$ cm) | $\rho_{yx}$ | Hall resistivity ($\Omega$ cm) |
| $\sigma_{xx}$ | Electrical conductivity (S cm$^{-1}$) | $\sigma_{xy}$ | Hall conductivity (S cm$^{-1}$) |
| $\kappa_{xx}$ | Thermal conductivity (W m$^{-1}$ K$^{-1}$) | | |

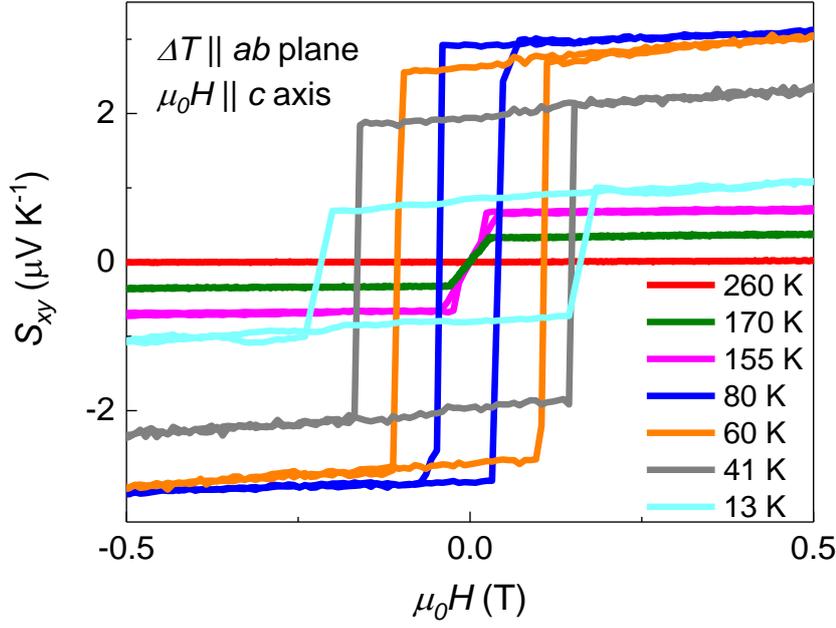

**Figure S4.** Magnetic-field-dependent Nernst thermopower ($S_{xy}$) data discussed in the main text at selected temperatures up to room temperature.

**Magnetization (*M*) dependence of the anomalous Nernst thermopower ($S_{xy}^A$).** The ANE was observed for the first time in topologically trivial ferromagnets. Therefore, it was believed for a long time that, as for the AHE, the ANE is proportional to the magnetization of the material. The linear scaling relation for conventional ferromagnetic metals can be expressed as: $|S_{xy}^A| = |Q_{xy}^A|\mu_0 M$, where $S_{xy}^A$ is the anomalous Nernst thermopower, $\mu_0 M$ is the magnetization (in Tesla), and $Q_{xy}^A$ is the anomalous Nernst coefficient. For topologically trivial ferromagnets, $Q_{xy}^A$ is in the range of ∼0.05 to ∼1 µV K$^{-1}$ T$^{-1}$. More detailed discussions are presented in refs. [S5,S6]. By converting the magnetization in Tesla and using the conventional scaling relation, we expect a maximum $S_{xy}^A$ of Co$_3$Sn$_2$S$_2$ of ~4.25 × 10$^{-3}$ to 0.0854 µV K$^{-1}$, significantly lower than the measured value. Therefore, $S_{xy}^A$ of Co$_3$Sn$_2$S$_2$ does not follow the conventional scaling relation, which indicates that the origin can be attributed to the topological band structure effect.

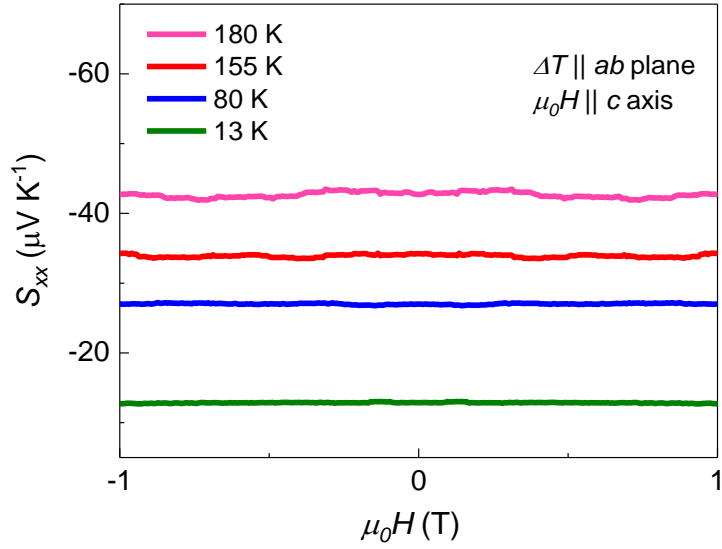

**Figure S5.** Magneto-Seebeck thermopower ($S_{xx}$) of $Co_3Sn_2S_2$. $S_{xx}$ almost does not exhibit field dependence at all of the measurement temperatures.

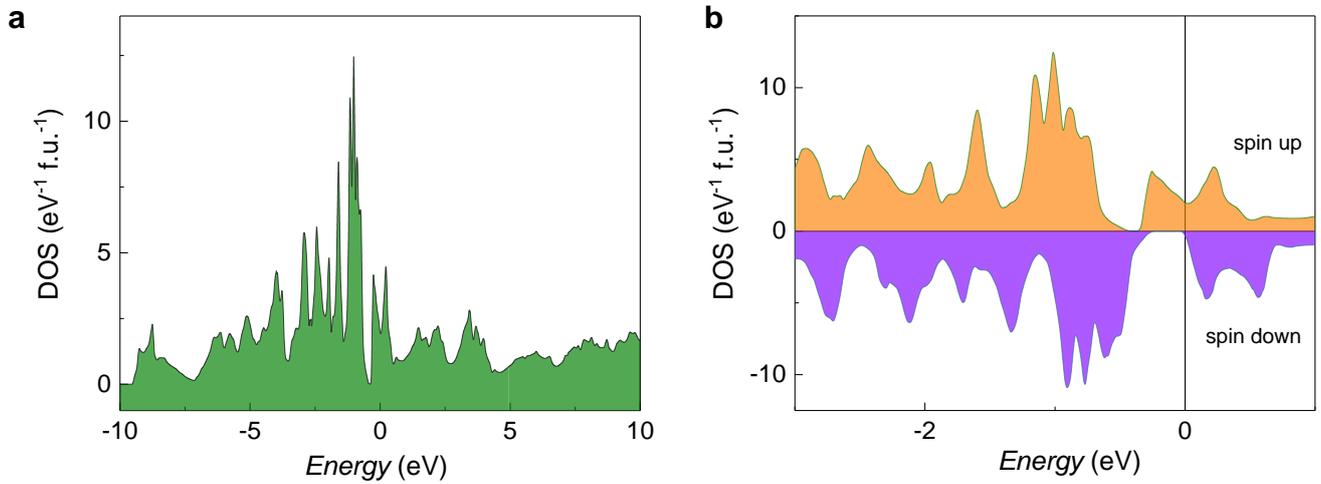

**Figure S6.** Electronic density of states (DOS) of $Co_3Sn_2S_2$. a) Total non-spin-resolved and b) spin-resolved DOSs. The spin-down channel has a gap near the Fermi level, whereas the spin-up channel is semimetallic.

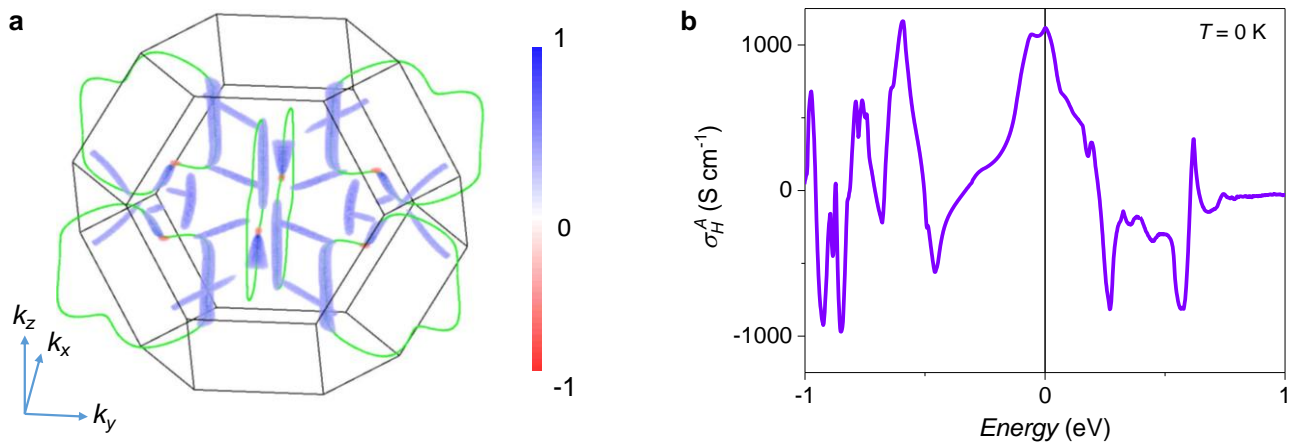

**Figure S7.** Berry curvature distribution and Hall conductivity ($\sigma_H^A$) in the Hall effect. a) Berry curvature distribution for $\sigma_H^A$ in the Brillouin zone with the magnetization along the *c*-direction at $T = 0$ K. b) Theoretically estimated $\sigma_H^A$ as a function of the chemical potential at $T = 0$ K.

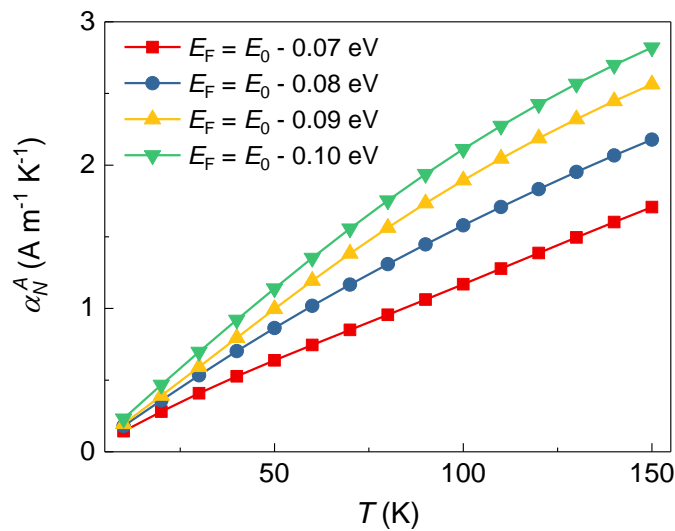

**Figure S8.** Temperature dependence of $\alpha_N^A$ below ordering temperature for $E_F = E_0 - 0.07$ eV to $E_0 - 0.10$ eV.

**Discussion on finite temperature DFT calculation.**

It is well known that the DFT only works for the ground state at 0 K. However, we can estimate lots of useful information at finite temperature from the ground state band structure if the band structure effect is robust enough. $Co_3Sn_2S_2$ possess a robust topological band structure and the transverse transport properties are mainly dominated by the intrinsic contribution from Berry curvature effect. The robust of Berry curvature effect can be seen from temperature dependent anomalous Hall measurements (Figure 3c).[S1] The anomalous Hall conductivity almost keeps at a constant value of 1100 S/cm from 2 K to 100 K, which is very close to the calculated value at 0 K. This good agreement should not be accidental. After the detailed analysis of Berry curvature, we think the origin of this good agreement is due to robust topological band structure. For anomalous Nernst effect, there is also an explicit logic for the Berry phase contribution, which mainly dependents of ground state band structure and temperature effect, as presented in the method part of the main manuscript.[S7] For the spin fluctuations, we already paid attention in our previous works.[S1] We have tried some possible high symmetry non-collinear magnetic structures in first principle calculations and found that the linear magnetic structure gives the lowest total energy and best fit to the experimental results, such as net magnetic moments, anomalous Hall conductivity, and ARPES band structure fitting. In addition, we also checked the robust of the electronic band structure with the non-collinear band structure, where the topological feature of Weyl points is very stable with non-collinear magnetic structure perturbations.[S1]

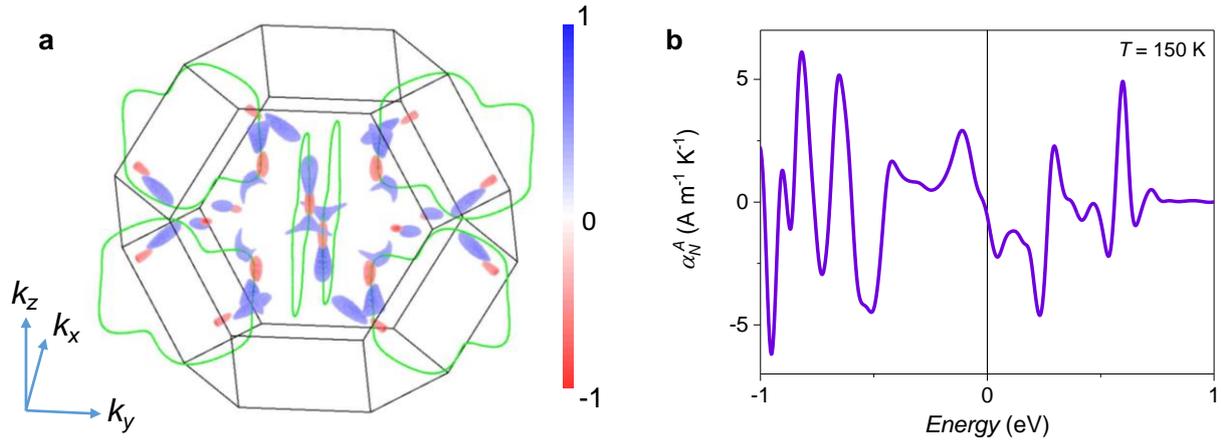

**Figure S9.** Berry curvature distribution and ANC ($\alpha_N^A$) in the Nernst effect. a) Berry curvature distribution in the Brillouin zone with the magnetization along the *c*-direction for $\alpha_N^A$ at $T = 150$ K for $E_F = E_0 - 0.08$ eV. b) Theoretically estimated $\alpha_N^A$ as a function of the chemical potential at $T = 150$ K.

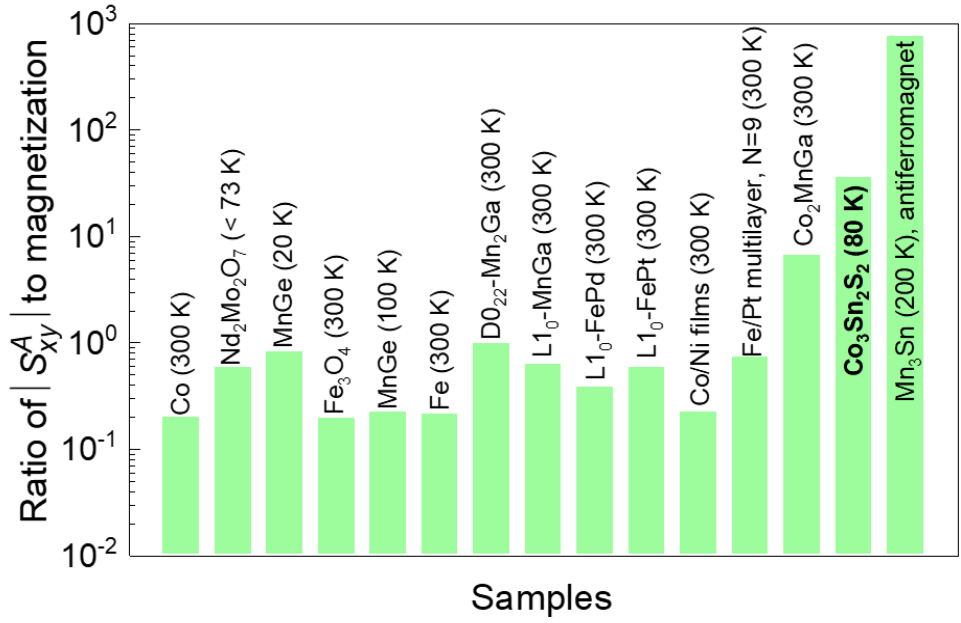

**Figure S10.** Comparison of ratio of anomalous Nernst thermopower ($S_{xy}^A$) to the magnetization ($\mu_0 M$) of various ferromagnetic metals, antiferromaget $Mn_3Sn$ and $Co_3Sn_2S_2$ (see Methods for details).

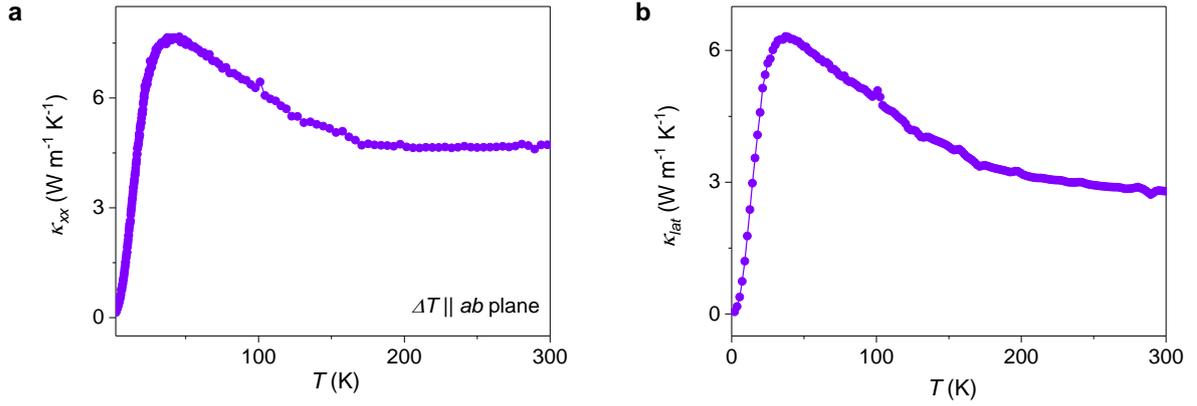

**Figure S11.** Thermal conductivity of $Co_3Sn_2S_2$. a) Temperature dependence of the total thermal conductivity ($\kappa_{xx}$) of $Co_3Sn_2S_2$. b) Temperature dependence of the lattice thermal conductivity ($\kappa_{lat}$) extracted after subtraction of $\kappa_{carrier}$.

$\kappa_{xx}$ of the crystal in the *ab*-plane is very low considering the metallic nature of the material. We obtained the temperature dependence of the lattice contribution $\kappa_{lat}$ by subtracting $\kappa_{carrier}$ using the Wiedemann–Franz relation, $\kappa_{carrier} = L\sigma_{xx}T$, where $L$ is the Lorenz number ($2.45 \times 10^{-8}$ W $\Omega$ $K^{-2}$), $\sigma_{xx}$ is the electrical conductivity, and $T$ is the temperature.